\documentstyle[11pt,epsfig]{article}
\title{\bf Classical and quantum dynamics of confined test particles in brane gravity}
\author{S. Jalalzadeh$^1$\thanks{email: s-jalalzadeh@sbu.ir}
  and H. R. Sepangi$^{1,2}$\thanks{email:
hr-sepangi@sbu.ir}
\\ $^1${\small Department of Physics, Shahid Beheshti University, Evin, Tehran 19839, Iran}\\$^2${\small
Institute for Studies in Theoretical Physics and Mathematics, P.O.
Box 19395-5746, Tehran, Iran }}
\begin{document}
\maketitle 
\begin{abstract}
  A model is constructed for the confinement of  test particles moving on a
  brane. Within the classical framework of this
  theory, confining a test particle to the brane eliminates the
  effects of extra dimensions, rendering them undetectable.
  However, in the quantized version of the theory, the effects of the gauge
  fields and extrinsic curvature are pronounced and this might
  provide  a hint for detecting them. As a  consequence of confinement the mass of the
  test particle is shown to be quantized. The condition of stability against
  small perturbations   along extra dimensions is also studied and its relation
  to dark matter is discussed. \vspace{5mm}
\end{abstract}\pagebreak
\section{Introduction}
The first exploration of the idea of extra dimensions was made by
Kaluza and Klien (KK). In this theory the gravitational and
electromagnetic interactions have a common origin. This
unification and its generalization to  Yang-Mills interactions can
only be done in the presence of extra dimensions.  Another reason
to study extra coordinates comes from string theory which is a
candidate theory for quantum gravity, being formulated
consistently in spaces with extra dimensions. The third reason to
study extra dimensions arises  from cosmological constant problem
(CCP). This is one of the most sever problems facing gravitational
and particle physics. The CCP can be put in two categories. The
first is the question of why the observed value of the vacuum
energy density is so small that the ratio of its experimental to
the theoretical value is of the order of -120. The other is to
understand why the observed vacuum energy density is not only
small, but also, as current Type Ia supernova observations seem to
indicate, of the same order of magnitude as the present mass
density of the universe \cite{shen}.

If the extra dimensions exist, then the natural
 question that one might ask is: why can we not travel through them and by which
 mechanism are they  detectable?
 In KK theory the extra dimensions have compact
 topology with certain compactification scale $l$ so that at the
 scales much larger than $l$, the extra dimensions should
 not be observable. They only become visible when one probes at very
 short distances of order $l$.
 Another way to display the invisible nature of
 extra dimensions in low energy scales is to assume that the
 standard matter is confined to a 4-dimensional submanifold
 (brane) embedded in a higher dimensional manifold (bulk), while
 the extra dimensions are probed by gravitons.
 An example comes from Horava-Witten's M-theory \cite{horava}, where the
 standard model of interactions contained in the
 $E_{8}\times E_{8}$ heterotic string theory is also confined to a 3-brane,
 but gravitons propagate in the 11-dimensional bulk.

In this paper, we have studied the dynamics of  test particles
confined to a brane at classical and quantum levels. In doing so,
we have assumed an $m$-dimensional bulk space through which a 4$D$
brane can move. The existence of more than one extra dimension and
certain Killing vector fields suggest that the twisting vector
fields, as in classical Kaluza-Klein theory, play the role of the
gauge fields. We then move on to study the classical dynamics of a
confined test particle. This in turn requires an algebraic
constraint being imposed on the extrinsic curvature. An
interesting question would be to investigate the effects of small
perturbations along the extra dimensions.  It turns out that
within the classical limits, the particle remains stable under
small perturbations and the effects of extra dimensions are not
felt by the test particle, hence making them  undetectable in this
way. At the quantum level however, since the uncertainty principle
prevents the wavefunction of a test particle from being exactly
localized on the brane,  the gauge fields and extrinsic curvature
effects become pronounced in the Klein-Gordon equation induced on
the brane. Ultimately, the quantum fluctuations of the brane cause
the mass of a test particle to become quantized. The cosmological
constant problem is also addressed within the context of this
approach. We show that the difference between the values of the
cosmological constant in particle physics and cosmology stems from
our measurements in two different scales, small and large.

\section{Geometrical preliminary of  the model}
Consider the background manifold $ \overline{V}_{4} $
isometrically embedded in a pseudo-Riemannian manifold $ V_{m}$ by
the map ${ \cal Y} : \overline{V}_{4}\rightarrow  V_{m} $ such
that
\begin{eqnarray}
{\cal G} _{AB} {\cal Y}^{A}_{,\mu } {\cal Y}^{B}_{,\nu}=
\bar{g}_{\mu \nu}  , \hspace{.5 cm} {\cal G}_{AB}{\cal
Y}^{A}_{,\mu}{\cal N}^{B}_{a} = 0  ,\hspace{.5 cm}  {\cal
G}_{AB}{\cal N}^{A}_{a}{\cal N}^{B}_{b} = g_{ab}\equiv\epsilon_{a}
;\hspace{.5 cm} \epsilon_{a} = \pm 1
\end{eqnarray}
where $ {\cal G}_{AB} $  $ ( \bar{g}_{\mu\nu} ) $ is the metric of
the bulk (brane) space  $  V_{m}  (\overline{V}_{4}) $ in
arbitrary coordinates, $ \{ {\cal Y}^{A} \} $   $  (\{ x^{\mu} \})
$  is the  basis of the bulk (brane) and  ${\cal N}^{A}_{a}$ are
$(m-4)$ normal unite vectors, orthogonal to the brane. The
perturbation of $\bar{V}_{4}$ in a sufficiently small neighborhood
of the brane along an arbitrary transverse direction $\zeta$ is
given by
\begin{eqnarray}
{\cal Z}^{A}(x^{\mu},\xi^{a}) = {\cal Y}^{A} +
(\pounds_{\zeta}{\cal Y})^{A}, \label{eq}
\end{eqnarray}
where $\pounds$ represents the Lie derivative. By choosing
$\zeta^{a}$ orthogonal to the brane, we ensure gauge independency
\cite{maia1} and  have perturbations of the embedding along a
single orthogonal extra direction $\bar{{\cal N}}_{a}$ giving
local coordinates of the perturbed brane as
\begin{eqnarray}
{\cal Z}^{A}_{,\mu}(x^{\nu},\xi^{a}) = {\cal Y}^{A}_{,\mu} +
\xi^{a}\bar{{\cal N}}^{A}_{a,\mu}(x^{\nu}),
\end{eqnarray}
where $\xi^{a}$ $(a = 0,1,...,m-5)$ is a small parameter along
${\cal N}^{A}_{a}$ that parameterizes the extra noncompact
dimensions. Also one can see from equation (\ref{eq}) that since
the vectors $\bar{{\cal N}}^{A}$ depend only on the local
coordinates $x^{\mu}$, they do not propagate along the extra
dimensions
\begin{eqnarray}
{\cal N}^{A}_{a}(x^{\mu}) = \bar{{\cal N}}^{A}_{a} +
\xi^{b}[\bar{{\cal N}}_{b} , \bar{{\cal N}}_{a}]^{A} = \bar{{\cal
N}}^{A}_{a}.
\end{eqnarray}
The above  assumptions lead to the embedding equations of the
perturbed geometry
\begin{eqnarray}
g_{\mu\nu} = {\cal G}_{AB}{\cal Z}^{A}_{,\mu}{\cal Z}^{B}_{,\nu},
\hspace{.5 cm}
 g_{\mu a}={\cal G}_{AB}{\cal Z}^{A}_{,\mu}{\cal N}^{B}_{a},\hspace{.5 cm}
 {\cal G}_{AB}{\cal N}^{A}_{a}{\cal N}^{B}_{b} = g_{ab}.
 \end{eqnarray}
If we set ${\cal N}^{A}_{a} = \delta^{A}_{a}$, the metric of the
bulk space can  be written in the matrix form (Gaussian frame)
\begin{eqnarray}
{\cal G}_{AB} = \left(\!\!\!\begin{array}{cc}
  g_{\mu\nu} + A_{\mu c}A^{c}_{\nu} & A_{\mu a} \\
  A_{\nu b} & g_{ab}
\end{array}\!\!\!\right), \label{eq6}
\end{eqnarray}
where
\begin{eqnarray}
g_{\mu\nu} = \bar{g}_{\mu\nu} - 2\xi^{a}\bar{K}_{\mu\nu a} +
\xi^{a}\xi^{b}\bar{g}^{\alpha\beta}\bar{K}_{\mu\alpha
a}\bar{K}_{\nu\beta b}, \label{eq7}
\end{eqnarray}
is the metric of the perturbed brane, so that
\begin{eqnarray}
\bar{K}_{\mu\nu a} = -{\cal G}_{AB}{\cal Y}^{A}_{,\mu}{\cal
N}^{B}_{a;\nu},
\end{eqnarray}
represents the extrinsic curvature of the original brane (second
fundamental form). Also, we use the notation $A_{\mu c} =
\xi^{d}A_{\mu cd}$ where
\begin{eqnarray}
A_{\mu cd} = {\cal G}_{AB}{\cal N}^{A}_{\,\,\,\,d;\mu}{\cal
N}^{B}_{c} = \bar{A}_{\mu cd},
\end{eqnarray}
represents the twisting vector field (normal fundamental form).
Any fixed $\xi^{a}$ shows a new brane, enabling us to define an
extrinsic curvature similar to the original one by
\begin{eqnarray}
K_{\mu\nu a} = -{\cal G}_{AB}{\cal Z}^{A}_{,\mu}{\cal
N}^{B}_{a;\nu} = \bar{K}_{\mu\nu a} - \xi^{b}( \bar{K}_{\mu\gamma
a}\bar{K}^{\gamma}_{\,\,\,\,\nu b} + A_{\mu c
a}A^{c}_{\,\,\,\,b\mu} ). \label{eq10}
\end{eqnarray}
Note that definitions (\ref{eq7}) and (\ref{eq10}) require
\begin{eqnarray}
K_{\mu\nu a} = -\frac{1}{2}\frac{\partial {\cal
G}_{\mu\nu}}{\partial \xi^{a}},
\end{eqnarray}
which is the generalized York's relation and shows how the
extrinsic curvature propagates as a result of the propagation of
metric in the direction of extra dimensions. In general the new
submanifold is an embedding in such a way that the geometry and
topology of  the bulk space do not become fixed \cite{maia1}. We
now show that if the bulk space has  certain Killing vector
fields, then $A_{\mu a b}$ transform as the components of a gauge
vector field under the group of isometries of the bulk space.
Under a local infinitesimal coordinate transformation for extra
dimensions we have
\begin{eqnarray}
\xi'^a = \xi^{a} + \eta^{a}.
\end{eqnarray}
Assuming the coordinates of the brane are  fixed $x'^{\mu} =
x^{\mu}$ and defining
\begin{eqnarray}
\eta^{a} = {\cal M}^{a}_{\,\,\,b}\xi^{b},
\end{eqnarray}
then in the Gaussian coordinates of the bulk space (\ref{eq6}) we
have
\begin{eqnarray}
g'_{\mu a} = g_{\mu a} + g_{\mu b}\eta^{b}_{,a} +
g_{ba}\eta^{b}_{,\mu} + \eta^{b}g_{\mu a,b} + {\cal O}(\xi^2),
\end{eqnarray}
hence the transformation of $A_{\mu a b}$ becomes
\begin{eqnarray}
A'_{\mu a b} = \frac{\partial g'_{\mu a}}{\partial \xi'^{b}} =
\frac{\partial g'_{\mu a}}{\partial \xi^{b}} -
\eta^{A}_{,b}\frac{\partial g'_{\mu a}}{\partial x^{A}}.
\end{eqnarray}
Now, using $\eta^{a}_{,b}={\cal M}^{a}_{\,\,\,b}(x^{\mu})$ and
$\eta^{a}_{,\mu} = {\cal M}^{a}_{\,\,\,\,b,\mu}\xi^{b}$ we obtain
\begin{eqnarray}
A'_{\mu a b} = A_{\mu a b} - 2A_{\mu c[a}{\cal M}_{\,\,\,b]} ^{
 c} + {\cal M}_{ab,\mu}.
\end{eqnarray}
This is exactly the gauge transformation of a Yang-Mills gauge
potential. In our model the gauge potential can only be present if
the dimension of the bulk space is equal to or greater than six
$(m\geq 6)$, because the gauge fields $A_{\mu ab}$ are
antisymmetric under the exchange of extra coordinate indices $a$
and $b$. For example, let the bulk space has an isometry group
$SO(p-1,q-3)$. Then if ${\cal L}^{ab}$ is the Lie algebra
generators of the this group, we have
\begin{eqnarray}
[ {\cal L}^{ab},{\cal L}^{cd}] = C_{pq}^{abcd} {\cal L}^{pq},
\nonumber
\end{eqnarray}
where $C_{pq}^{abcd}$ is the Lie algebra structure constants
defined by
\begin{eqnarray}
C_{pq}^{abcd} = 2\delta_{p}^{[b}g^{a][c}\delta^{d]}_{q}.
\end{eqnarray}
On the other hand if $F^{\mu\nu} =
F^{\mu\nu}_{\,\,\,\,\,\,\,\,ab}{\cal L}^{ab}$ is to be the
curvature associated with the vector potential $A^{\mu} =
A^{\mu}_{\,\,\,\,ab}{\cal L}^{ab}$,  we have
\begin{eqnarray}
F_{\mu\nu} = A_{\nu,\mu} - A_{\mu,\nu} +
\frac{1}{2}[A_{\mu},A_{\nu}],
\end{eqnarray}
or in component form
\begin{eqnarray}
F_{\mu\nu ab} = A_{\nu, \mu ab} - A_{\mu, \nu ab} +
\frac{1}{2}C^{mnpq}_{ab}A_{\mu mn}A_{\nu pq}.
\end{eqnarray}
\section{Classical dynamics of test particles and confinement}
In this section, we apply the above formalism to derive the $4D$
geodesic equation for a particle confined to a brane and the
conditions for the confinement. The geodesic equation for a test
particle traveling in the neighborhood of  the brane world in
$V_{m}$ are taken to be
\begin{eqnarray}
\frac{d{\cal U}^{A}}{d\lambda} + \bar{\Gamma}^{A}_{BC}{\cal
U}^{B}{\cal U}^{C} = \frac{{\cal F}^{A}}{M^2}, \label{eq21}
\end{eqnarray}
where ${\cal U}^{A} = \frac{d{\cal Z}^{A}}{d\lambda}$, ${\cal
G}_{AB}{\cal U}^{A}{\cal U}^{B} = \kappa$, $M$ is the mass of test
particle in the bulk space, $\lambda$ is an affine parameter and
${\cal F}^{A}$ is some non-gravitational force that is responsible
for the confinement of test particles and defined by the potential
$\cal V$ such that ${\cal F}^a=-\nabla^a {\cal V}$. We require
${\cal F}^{A}$ to satisfy three general conditions: firstly, its
defining potential has a deep minimum on the non-perturbed brane,
secondly, depends only on extra coordinates and thirdly, the gauge
group representing the subgroup of the isometry group is preserved
by the potential. Here, $\kappa = -1,0,1$ to allow for massive,
null and tachyonic particles respectively. One can decompose these
equations by using the Guassian form of the bulk space metric. In
this frame the Christoffel symbols of $V_{m}$ can be written as
\begin{eqnarray}
\begin{array}{lllll}\vspace{.5 cm}
\bar{\Gamma}^{\mu}_{\alpha\beta} = \Gamma^{\mu}_{\alpha\beta} +
\frac{1}{2}\{ A_{\alpha c}F_{\beta}^{\,\,\,\mu c} + A_{\beta
c}F_{\alpha}^{\,\,\,\mu c} \} - K_{\alpha\beta a}A^{\mu a},\\
\vspace{.5 cm} \bar{\Gamma}^{\mu}_{\alpha a} =
-K^{\mu}_{\,\,\,\,\alpha a} - \frac{1}{2}F^{\mu}_{\,\,\,\,\alpha a},\\
\vspace{.5 cm} \bar{\Gamma}^{\mu}_{ab} = \bar{\Gamma}^{a}_{bc} =
0,\\ \vspace{.5 cm} \bar{\Gamma}^{a}_{\alpha\beta} =
\frac{1}{2}\left\{ \nabla_{\beta}A^{a}_{\,\,\,\,\alpha} +
\nabla_{\alpha}A^{a}_{\,\,\,\,\beta} +
A^{a\mu}A_{\,\,\,\,\alpha}^{c}F_{c\mu\beta} +
A^{a\mu}A_{\,\,\,\,\beta}^{c}F_{c\mu\alpha}\right\} +
K_{\,\,\,\,\alpha\beta}^{a} + A^{a}_{\,\,\,\,\mu}A^{\mu b}K_{\alpha\beta b},\\
\vspace{.5 cm} \bar{\Gamma}^{a}_{b \alpha} =
-\frac{1}{2}A^{a\beta}F_{b\alpha\beta} + A^{a}_{\,\,\,\,b\alpha},
\end{array}
\end{eqnarray}
where $\Gamma^{\mu}_{\alpha\beta}$ is the christoffel symbol
induced on the perturbed brane. Note that for obtaining these
relations we have used the inverse of ${\cal G}_{AB}$
\begin{eqnarray}
{\cal G}^{AB} = \left(\!\!\!\begin{array}{cc}
  g^{\mu\nu}  & -A^{\mu a} \\
 - A^{\nu b} & g^{ab} + A^{a}_{\alpha}A^{\alpha b}
\end{array}\!\!\!\right),
\end{eqnarray}
hence the geodesic equations (\ref{eq21}) in  the Guassian frame
splits into the following equations
\begin{eqnarray}
\frac{du^{\mu}}{d\lambda} + (\Gamma^{\mu}_{\alpha\beta} +
A_{\alpha c}F_{\beta}^{\,\,\,\,\mu c} - K_{\alpha\beta a}A^{\mu a}
)u^{\alpha}u^{\beta} - ( 2K^{\mu}_{\,\,\,\,\alpha a} +
F^{\mu}_{\,\,\,\,\alpha a} )u^{a}u^{\alpha} = \frac{{\cal
F}^{\mu}}{M^2}, \label{eq24}
\end{eqnarray}
\begin{eqnarray}
\frac{du^{a}}{d\lambda} + ( \nabla_{\alpha}A^{a}_{\,\,\,\,\beta} +
A^{a \mu}A_{\,\,\,\,\alpha}^{c}F_{c\mu \beta} +
K^{a}_{\,\,\,\,\alpha\beta} + A^{a}_{\,\,\,\,\mu}A^{\mu
b}K_{\alpha\beta b} )u^{\alpha}u^{\beta} - ( 2A_{\,\,\,\,\alpha
b}^{a} + A^{a\beta}F_{b \alpha\beta} )u^{\alpha}u^{b} =\frac{
{\cal F}^{a}}{M^2}, \label{eq25}
\end{eqnarray}
and the normalization condition ${\cal G}_{AB}{\cal U}^{A}{\cal
U}^{B} =\kappa$ becomes
\begin{eqnarray}
g_{\alpha\beta}u^{\alpha}u^{\beta} + g_{ab}( u^{a} +
A^{a}_{\,\,\,\,\mu}u^{\mu} )( u^{b} + A^{a}_{\,\,\,\,\nu}u^{\nu} )
= \kappa, \label{eq26}
\end{eqnarray}
where $u^{\mu} = \frac{dx^{\mu}}{d\lambda}$ and  $u^{a} =
\frac{d\xi^{a}}{d\lambda}$. Now, if a test particle is confined to
the original non-perturbed brane through the action of the force
$\cal F$ and if we impose the first of the constraining condition
which implies that ${\cal F}^{a}$ can be expanded as a power
series in $\xi^{a}$ about the minimum of the corresponding
potential, that is
\begin{eqnarray}
{\cal F}^{a} =-\omega^a \omega_b\xi^b+{\cal O}(\xi^2),
\end{eqnarray}
then the components of the force vanish on the brane with $u^{a} =
0 = \xi^{a}$. This is so because according to our assumptions the
potential is not a function of the brane coordinates $x^\mu$ and
therefore ${\cal F}^\mu=0$. Also ${\cal F}^a=0$ since the
potential has a minimum on the brane. Equations (\ref{eq24}),
(\ref{eq25}) and (\ref{eq26}) thus reduce to
\begin{eqnarray}
\frac{d\bar{u}^{\mu}}{ds} +
\Gamma^{\mu}_{\alpha\beta}\bar{u}^{\alpha}\bar{u}^{\beta} &=&0 , \\
\frac{d\bar{u}^{a}}{ds} + \bar{K}^{a}_{\,\,\,\,\alpha
\beta}\bar{u}^{\alpha}\bar{u}^{\beta}& =&0,
\label{eq27}\\
\bar{u}^{\mu}\bar{u}_{\mu}& =& \kappa,
\end{eqnarray}
where $\bar{u}^{\alpha} = dx^{\alpha}/ds$, $\bar{u}^{a} =
dx^{a}/ds$ and $s$ is an affine parameter on the brane. Here, in
the spirit of our third assumption about the general behavior of
the potential we have assumed that $\cal V$ preserves the
symmetries of the gauge fields. However, for later convenience we
assume that our gauge symmetry is that of the group $SO(p-1,q-3)$
and hence we may write $\cal V$ in a symmetric form given by
${\cal V}=\frac{1}{2}\omega^2 g_{ab}\xi^a\xi^b$. Since ${\cal V}$
has a deep minimum, we can neglect $\xi^2$ and its higher orders
when expanding the force. To cancel these terms, we consider
$\omega$ to be much larger than the inverse of the scale of
curvatures $\rho^{-1}$ on $V_4$, or more specifically $\omega\gg
\rho^{-2}$. Following \cite{jaffe}, we adsorb the scale of
$\omega$ into a small dimensionless parameter $\varepsilon$, that
is $\omega\rightarrow \omega/\varepsilon$, so that $\omega$
becomes of the same order as $\rho^{-2}$. The smaller the
``squeezing'' parameter $\varepsilon$ the deeper is the minimum of
${\cal V}$ and the system is more squeezed on the original brane.
Thus $\varepsilon$ plays the role of a natural perturbation
parameter. For a confined particle, equation (\ref{eq27}) requires
that acceleration along the extra coordinate must vanish, {\it
i.e.} $\frac{d\bar{u}^{a}}{ds} =0$. This will occur if
\begin{eqnarray}
\bar{K}^{a}_{\,\,\,\,\alpha\beta}\bar{u}^{\alpha}\bar{u}^{\beta}
=0. \label{eq29}
\end{eqnarray}
This is an algebraic constraint equation that must hold for the
confinement to happen, in agreement with the results obtained in
5$D$ in \cite{seahra}. As a special case, equation (\ref{eq29}) is
satisfied when $\bar{K}^{a}_{\alpha\beta} =0$. In this case the
brane is a totaly geodesic sub-manifold, that is, any geodesic of
$V_{4}$ is also a geodesic of $V_{m}$. A totally geodesic
submanifold is a multidimensional analogue of a geodesic line. A
Riemannian manifold containing a totally geodesic submanifold
cannot be arbitrary. Ricci \cite{ricci} has given a system of
differential equations that a Reimannian submanifold has to
satisfy in order to admit  totally geodesic submanifolds.

The above discussion means that at the classical level a test
particle  does not feel the effects of extra dimensions at low
energy. Later, we show that at the quantum level, the effects of
extra dimensions would become manifest as the gauge fields and a
potential characterized by the extrinsic curvature.
\section{Stabilization}
At this point it would be interesting to consider what would
happen if the position of the particle was perturbed along a
direction normal to the brane. In other words, how stable the
particle is confined to the brane. However, before doing so, it
would be necessary to make some of the concepts to be used in what
follows more transparent and clear. Let us then start by making a
quick look at the 5-dimensional brane world scenario  according to
the formulation presented in \cite{shiro}. This would help us
grasp the salient points of our discussion more easily.

In the SMS formalism, Einstein equations in the bulk space can be
written in the form
\begin{eqnarray}
^{(b)}G_{AB}=\kappa^2_{(5)}T_{AB}, \label{eqq1}
\end{eqnarray}
where $^{(b)}G_{AB}$ is the Einstein tensor in the bulk space,
$\kappa_{(5)}=1/M^3_{*}$, $M_{*}$ being the bulk energy scale and
\begin{eqnarray}
T_{AB}=-\Lambda_{(5)}{\cal G}_{AB}+\delta(\xi)S_{AB}. \label{eqq2}
\end{eqnarray}
Here, $S_{AB}$ is the energy-momentum tensor on the brane with
$S_{AB}{\cal N}^A=0$. This tensor consists of two part, that is
\begin{eqnarray}
S_{\mu\nu}=-\sigma g_{\mu\nu}+T_{\mu\nu}, \label{eqq3}
\end{eqnarray}
where $\sigma$ is the tension of the brane in $5D$ and
$T_{\mu\nu}$ is the energy-momentum tensor of ordinary matter on
the brane. We note that the existence of the $\delta$ function in
the energy-momentum tensor (\ref{eqq2}) leads to the usual
Israel-Lanczos junction conditions
\begin{eqnarray}
[g_{\mu\nu}]=0 \hspace{3mm}\mbox{and}\hspace{5mm} [K_{\mu\nu}]=
-\kappa^2_{(5)}(S_{\mu\nu}-\frac{1}{3}g_{\mu\nu}S) ,
\end{eqnarray}
where $[X]=\mbox{lim}_{\xi\rightarrow
0^{+}}X-\mbox{lim}_{\xi\rightarrow 0^{-}}X.$ Now, imposing $Z_2$
symmetry on the bulk space and considering the brane as fixed,
this symmetry determines the extrinsic curvature of the brane in
terms of the energy-momentum tensor
\begin{eqnarray}
K^{+}_{\mu\nu}=\frac{1}{2}\kappa^2_{(5)}\left(S_{\mu\nu}-
\frac{1}{3}g_{\mu\nu}S\right).
\end{eqnarray}

Now, to move any further we have to write the Einstein field
equations. There are two ways in which the brane world Einstein
field equations differ from what is customary. Firstly, they do
not constitute a closed system in that they contain an unspecified
electric part of the Weyl tensor, where it can only be specified
in terms of the bulk properties. Secondly, they contain a term
quadratic in the energy-momentum tensor of the brane $S_{\mu\nu}$,
which is important for the evolution of baby universe models
\cite{baby}. However, this is not the end of story. According to
\cite{tavakol}, the splitting of the right hand side of the
brane-Einstein equations into a term characterizing the bulk
(${\cal E}_{\mu\nu}$) and a term on the brane containing linear
and quadratic terms of the energy-momentum tensor is highly
non-unique. Since in the present model we do not restrict
ourselves to one extra dimension, there will be no suitable
Israel-Lanczos junction conditions and so the usual handling of
the brane-Einstein equations would break down. The reason for this
is that if the number of extra dimensions exceeds one, the brane
cannot be considered as a boundary between two regions.

In the spirit of the above discussions and because of the fact
that we are using a confining potential approach rather than
matter localization by a delta-function in $T_{AB}$,  the
extrinsic curvature would be independent of the matter content of
the brane, in contrast to using the junction conditions. Thus the
arguments presented in reference \cite{tavakol} regarding the
appearing or vanishing of the quadratic terms of the energy
momentum tensor in Einstein field equations on the brane world
would be rendered unnecessary.  Let us then start by contracting
the Gauss-Codazzi relations \cite{eizen}
\begin{eqnarray}
R_{\alpha\beta\gamma\delta} &=&
2g^{\mu\nu}K_{\alpha[\gamma\mu}K_{\beta]\delta\nu} + {\cal
R}_{ABCD}{\cal Z}^{A}_{,\alpha}{\cal Z}^{B}_{,\beta}{\cal
Z}^{C}_{,\gamma}{\cal Z}^{D}_{,\delta}\nonumber\\
\label{eqq5}\\
 K_{\alpha[\gamma b;\delta]} &=& g^{mn}A_{[\gamma m
a}K_{\alpha \beta] n} + {\cal R}_{ABCD}{\cal Z}^{A}_{,\alpha}{\cal
N}^{B}_{b}{\cal Z}^{C}_{,\gamma}{\cal Z}^{D}_{,\delta},\nonumber
\end{eqnarray}
where ${\cal R}_{ABCD}$ and $R_{\alpha\beta\gamma\delta}$ are the
Riemann tensors for the bulk and the brane respectively, one
obtains
\begin{eqnarray}
^{(b)}G_{AB}{\cal Z}^A_{,\mu}{\cal
Z}^B_{,\nu}=G_{\mu\nu}-Q_{\mu\nu}-g^{ab}{\cal R}_{AB}{\cal N}^A_a
{\cal N}^B_b g_{\mu\nu}+g^{ab}{\cal R}_{ABCD}{\cal N}^A_a {\cal
Z}^B_{,\mu}{\cal Z}^C_{,\nu}{\cal N}^D_b, \label{eqq6}
\end{eqnarray}
where $G_{\mu\nu}$ is the Einstein tensor of the brane and
\begin{eqnarray}
Q_{\mu\nu}=g^{ab}\left(K^\gamma_{\mu a}K_{\gamma\nu b}-K_a
K_{\mu\nu b}\right)-\frac{1}{2}\left(K_{\alpha\beta
a}K^{\alpha\beta a}-K_a K^a\right)g_{\mu\nu}, \label{eqq7}
\end{eqnarray}
with $K_a=g^{\mu\nu}K_{\mu\nu a}$. Note that directly from the
definition of $Q_{\mu\nu}$, it follows that it is independently a
conserved quantity, that is $Q^{\mu\nu}_{\,\,\,\,\,\,\,;\mu}=0$.

Now, in order to substitute for the terms proportional to the bulk
space in equation (\ref{eqq6}), we use Einstein equation
\begin{eqnarray}
^{(b)}G_{AB}+^{(b)}\!\!\Lambda{\cal G}_{AB}=\alpha^{*}S_{AB},
\label{eqqq8}
\end{eqnarray}
where $\alpha^{*}=1/M^{m-2}_{*}$. Also, $^{(b)}\!\Lambda$ is the
cosmological constant of the bulk space with $S_{AB}$ consisting
of two parts
\begin{eqnarray}
S_{AB}=T_{AB}+\frac{1}{2}{\cal V}{\cal G}_{AB}, \label{eqq9}
\end{eqnarray}
where $T_{AB}$ is the energy-momentum tensor of the matter
confining to the brane through the action of the confining
potential $\cal V$. The contracted Bianchi identities
$^{(b)}G^{AB}_{\,\,\,\,\, ;A}=0$ imply
\begin{eqnarray}
\left(T^{AB}+\frac{1}{2}{\cal V}{\cal G}^{AB}\right)_{; A}=0.
\label{qq}
\end{eqnarray}
If we take $T^{AB}$ as the energy-momentum tensor of the test
particle, the equations of motion (\ref{eq21}) are obtained. Also,
since there is no singular behavior in $T^{AB}$, we obtain the
following junction condition
\begin{eqnarray}
[K_{\mu\nu}]=0.
\end{eqnarray}
The application of  $Z_2$ symmetry now causes the brane to become
simply totally geodesic, that is $K^{+}_{\mu\nu}=0$. It should be
mentioned at this point that we do not apply such a symmetry to
our model since doing so would impose restrictions on the type of
the bulk space one may wish to consider.

Now, by decomposing the Riemann tensor into the Weyl and Ricci
tensors and Ricci
scalar respectively, we finally obtain
\begin{eqnarray}
G_{\mu\nu}&=&\alpha\tau_{\mu\nu}-\frac{m-4}{(m-1)(m-2)}\alpha\tau
g_{\mu\nu}\nonumber\\
&-&\left[\frac{(m-1)(m-2)^2+2(m-4)+(m-4)(m-1)^2(m-2)}{(m-1)(m-2)^2}\right]
{^{(b)}}\Lambda g_{\mu\nu}+Q_{\mu\nu}+{\cal E}_{\mu\nu},
\label{111}
\end{eqnarray}
where ${\cal E}_{\mu\nu}=g^{ab}{\cal C}_{ABCD}{\cal N}^A_a{\cal
Y}^B_\mu{\cal N}^C_b{\cal Y}^D_\nu$ and following \cite{maia1} we
find, $\alpha^{*}\sim (\alpha+1/M^2_e)V$. Here
$$\alpha=1/M^2_{Pl}, \hspace{3mm} 1/M^2_e=\int(K^{\mu\nu
a}K_{\mu\nu a}+K^a K_a)\sqrt{g}d^4x$$ and $V$ is the volume of the
extra space. As was mentioned before, $Q_{\mu\nu}$ is a conserved
quantity which according to \cite{maia2} may be considered as an
energy-momentum tensor of a dark energy fluid representing the
$x$-matter, the more common phrase is ``X-Cold-Dark Matter''
(XCDM). This matter has the most general form of the equation of
state which is characterized by the following conditions \cite{x}:
first it violates the strong energy condition at the present epoch
for $\omega_x<-1/3$ where $p_x=\omega_x\rho_x$, second it is
locally stable {\it i.e.} $c^2_s=\delta p_x/\delta\rho_x\ge 0$ and
third, the causality is granted, that is $c_s\le 1$. Ultimately,
we have three different types of ``matter,'' on the right hand
side of equation (\ref{111}) namely, ordinary confined conserved
matter (see equation (\ref{qq})) represented by $\tau_{\mu\nu}$,
the matter represented by $Q_{\mu\nu}$ which is independently
conserved and will be discussed in what follows and finally, the
Weyl matter represented by ${\cal E}_{\mu\nu}$. It follows that
${\cal E}^{\mu\nu}_{\,\,\,\,\,\,\,;\mu}\sim \tau^{,\nu}$.

It is now appropriate to go back to the task at hand and study the
stabilization of the confinement of our test particle. Since
$\xi^{a}$ is ``small'' in our approximation, equation (\ref{eq25})
up to ${\cal O}(\xi ^{2})$ becomes, calculating all the quantities
on the non-perturbed brane and disregarding the bar from hereon
\begin{eqnarray}
\frac{d^2\xi^{a}}{d\lambda^2} + ( A^{a}_{\,\,\,\,\beta;\alpha} +
K^{a}_{\,\,\,\,\alpha \beta} )u^{\alpha}u^{\beta} =
-\frac{\omega^2}{M^2\varepsilon^2}\xi^{a}.
\end{eqnarray}
Now, using equation (\ref{eq10}) and condition (\ref{eq29}) the
above equation simplifies to
\begin{eqnarray}
\frac{d^2\xi^{a}}{d\lambda^2}+\left[\left(A^a_{\,\,\,\,b\beta;\alpha}
-K^a_{\,\,\,\,\alpha\gamma}K^\gamma_{\,\,\,\,\beta
b}-A^a_{\,\,\,\,\alpha c}A^c_{\,\,\,\,b\beta}\right)u^\alpha
u^\beta+\frac{\omega^2}{M^2\varepsilon^2}\delta^a_{\,\,\,\,b}\right]\xi^b=0.
\label{eqq1}
\end{eqnarray}
For stabilizing the particle on the brane it would be necessary
for the term in the square brackets in equation (\ref{eqq1}) to be
positive. This suggests that we should write it in terms of the
quantities defined above such that our brane scenario acquires
meaningful interpretations.
 In an arbitrary m-dimensional bulk
space the Ricci equations are given by \cite{eizen}
\begin{eqnarray}
\label{eqq5}
 F_{\gamma\delta ab} &=&
-2g^{\mu\nu}K_{[\gamma \mu a}K_{\delta]\nu b} - {\cal
R}_{ABCD}{\cal N}^{A}_{a}{\cal N}^{B}_{b}{\cal
Y}^{C}_{,\gamma}{\cal Y}^{D}_{,\delta}.
\end{eqnarray}
Now, using the decomposition of the Riemann tensor in terms of the
Weyl tensor, one can show that
\begin{eqnarray}
{\cal R}_{ABCD}{\cal N}^{A}_{a}{\cal N}^{B}_{b}{\cal
Y}^{C}_{,\gamma}{\cal Y}^{D}_{,\delta}=0,
\end{eqnarray}
and hence the Ricci and Codazzi relations lead to
\begin{eqnarray}
F_{\mu\nu ab} = Q_{\nu\mu ab} - Q_{\mu\nu ab},
\end{eqnarray}
where
\begin{eqnarray}
Q_{\mu\nu ab} = K_{\mu\gamma a}K^{\gamma}_{\,\,\,\,\nu b} -
h_{a}K_{\mu\nu b} -
\frac{1}{2}g_{\mu\nu}(K^{\alpha\beta}_{\,\,\,\,\,\,\,\,a}K_{\alpha\beta
b} - h_{a}h_{b}). \label{eqq7}
\end{eqnarray}
The above considerations lead us to obtain the generalized XCDM
corresponding to an extrinsic quantity $Q_{\mu\nu ab}$ given by
\begin{eqnarray}
Q_{\mu\nu ab} = -8\pi G\left[(p_{x} + \rho_{x})u_{\mu}u_{\nu} +
p_{x}g_{\mu\nu}\right]g_{ab} + A_{\nu;\mu ab} - A_{\mu
ca}A_{\,\,\,\,\nu
 b}^{c}. \label{eqq8}
\end{eqnarray}
Multiplication of equations  (\ref{eqq7}) and (\ref{eqq8}) by
$u^\mu u^\nu$ results in
\begin{eqnarray}
\left(K_{\mu\gamma a}K^{\gamma}_{\,\,\,\nu b} - A_{\nu ; \mu ab} +
A_{\mu ca}A^{c}_{\,\,\,\nu b}\right)u^{\mu}u^{\nu} = -4\pi G
\left(3p_{x} + \rho_{x}\right)g_{ab}+\frac{1}{2}D_\mu
A^\mu_{\,\,\,\,ab}, \label{eqq9}
\end{eqnarray}
where $D_\nu A^\mu_{\,\,\,\,ab}=A^\mu_{\,\,\,\,ab;\nu}-A_{\nu
ac}{A^\mu_{\,\,\,\,b}}^c$. Hence if we assume that, as a
generalized Coulomb gauge $D_\nu A^\mu_{\,\,\,\,ab}=0$, then from
equation (\ref{eqq1}) and (\ref{eqq9}) we obtain
\begin{eqnarray}
\frac{\omega^2}{M^2\varepsilon^2} + 4\pi G(3\omega_{x}+1)\rho_{x}
> 0. \label{eqq10}
\end{eqnarray}
Since we know from accelerated expanding universe that
$\omega_x<-1/3$, the second term in equation (\ref{eqq10}) is
negative. However, since $\varepsilon\rightarrow 0$ the first term
is large and we should not worry about the particle being confined
to the brane.
\section{Quantization}
To describe the quantum dynamics of a test particle confined to a
semi-Riemannian submanifold embedded in a generic semi-Riemannian
manifold that satisfies Einstein field equations (\ref{eqqq8}),
three different approaches may be adapted. One is the confining
approach in which, as was discussed before, a confining potential
forces a particle to stay on the brane by a non-gravitational
force. Both of the other approaches follow Dirac's general
prescription to treat a constrained system, with two different
types of constraints, namely the usual and the conservative
constraints respectively.

Suppose that $f({\cal Y})=0$ defines our brane. The usual approach
employs this equation as the constraint condition and corresponds
to a dynamical system defined by the d'Alembert-Lagrange principle
\cite{seahra}. However one may use another approach in which the
constrain is given by $\frac{df}{ds}=0$ where $s$ is the affine
parameter on the brane \cite{mit}. This means that no dynamical
motion normal to the brane is allowed. The confining approach is a
rather straightforward one whereas there is much to be said about
the approach that follows Dirac's prescription. First, there is
the well known problem of the ordering of operators. Second, as
was mentioned above, the equations expressing the constraints are
not unique.  At the classical level the above three approaches
lead to the same classical equations of motion. However at the
quantum level, they generally lead to results which do not agree
with each other. In the rest of this section we concentrate on the
confining approach since realistically, in any real system the
transverse direction contains a number of atoms so any layer
cannot become smaller than $\hbar$ which is of order of the
magnitude of atomic dimensions, in units in which time and mass
are of order one. As regards our brane, the above discussion would
motivate us to use the confining approach since we assume that our
brane has a finite small thickness. It also suggests to link the
squeezing parameter, determined by the constraining potential, to
the quantum scale given by $\hbar$, that is
\begin{eqnarray}
\varepsilon=a\hbar^b, \label{eq1122}
\end{eqnarray}
where $a$ is a dimensional constants so as to making $\varepsilon$
dimensionless.

We now focus attention on the quantum aspects of the problem. The
Hamiltonian of the system in the coordinates of the bulk space is
\begin{eqnarray}
{\cal H} = \frac{1}{2}P_{A}P^{A} + {\cal V} - \frac{1}{2}M^2 =0.
\end{eqnarray}
Here, we have added to the Hamiltonian the confinement potential
$\cal V$, since by taking the covariant derivative of ${\cal H}$
we expect to obtain  equation (\ref{eq21}). Before considering the
constraint, the dynamics of the quantum particle is described by
the Klein-Gordon (KG) equation in the bulk space given by
\begin{eqnarray}
( -\frac{1}{2}{\cal G}^{AB}\nabla_{A}\nabla_{B} - \frac{1}{2}M^2
+{\cal V})\psi =0,
\end{eqnarray}
with the normalization condition for the wavefunction given by
\begin{eqnarray}
\int {|\psi|}^2d^{m}{\cal Z} = 1. \label{eq32}
\end{eqnarray}
By changing the coordinates to  $\{x^{\mu},\xi^{a}\}$, the
Dalambertian  in the KG equation becomes
\begin{eqnarray}
\triangle = -\frac{1}{2{|{\cal G}|}^{1/2}}\partial_{A}{\cal
G}^{AB}{|{\cal G}|}^{1/2}\partial_{B}
\end{eqnarray}
and the normalization condition (\ref{eq32}) takes the form
\begin{eqnarray}
\int {|\psi| }^2 {|{\cal G}|}^{1/2}dx^4d\xi^{m-4} =1.
\end{eqnarray}
Since our goal is to have an effective dynamics on the brane, we
re-scale  the wavefunction in such a way as to make it normalized
in $L^2(V_{4})$ instead of $L^2(V_{m})$. This aim is achieved by
the following re-scalings
\begin{eqnarray}
 \Phi &=& \frac{{|\bar{g}|}^{1/4}}{{|{\cal
G}|}^{1/4}} \psi \nonumber\\
 \Box &=&\frac{{|\bar{g}|}^{1/4}}{{|{\cal
G}|}^{1/4}}\triangle
 \frac{{|{\cal
G}|}^{1/4}}{{|\bar{g}|}^{1/4}},
 \end{eqnarray}
where ${\bar{g}}$ is the determinant of the non-perturbed metric
$\bar{g}_{\mu\nu}$. The re-scaled wavefunction then satisfies the
normalization condition
\begin{eqnarray}
\int d^4x d\xi^{m-4} {|\Phi|}^2 {|\bar{g}|}^{1/2} = 1.
\end{eqnarray}
Now, using the explicit form of the Gaussian metric we obtain
\begin{eqnarray}
 \Box&=&
-\frac{1}{2|g|^{1/4}}\partial_{a}|g|^{1/2}\partial^{a}\frac{1}{|g|^{1/4}}
- \frac{1}{2|\bar{g}|^{1/4}|g|^{1/4}} \left(
\partial_{\mu}g^{\mu\nu}|g|^{1/2}\partial_{\nu}\right. \nonumber\\
&+&\left. \xi^{a}\xi^{b}A^{m}_{\,\,\,\,\mu a}A^{n}_{\,\,\,\,\nu
b}\partial_{m}g^{\mu\nu}|g|^{1/2}\partial_{n} +
\partial_{\mu}g^{\mu\alpha}\xi^{a}A_{\,\,\,\,\alpha
a}^{b}|g|^{1/2}\partial_{b} +
\partial_{a}g^{\alpha\beta}\xi^{b}A_{\,\,\,\,\alpha
b}^{a}|g|^{1/2}\partial_{\beta}
\right)\frac{|\bar{g}|^{1/4}}{|g|^{1/4}}. \label{eq37}
\end{eqnarray}
Defining ${\cal D}_{\mu} = \partial_{\mu} - \frac{1}{2}iA_{\mu}$,
where $A_{\mu} = i A_{\mu a b}{\cal L}^{ab}$ and ${\cal L}^{ab}$
are the Lie algebra operators, the Dalambertian  can be rewritten
as
\begin{eqnarray}
\Box=
-\frac{1}{2|g|^{1/2}}\partial_{a}|g|^{1/2}\partial^{a}\frac{1}{|g|^{1/4}}
- \frac{1}{2|\bar{g}|^{1/4}|g|^{1/4}}{\cal
D}_{\mu}g^{\mu\nu}|g|^{1/4}{\cal
D}_{\nu}\frac{|\bar{g}|^{1/4}}{|g|^{1/4}}. \label{eq38}
\end{eqnarray}

In the previous section we saw that the parameter $\varepsilon$
can be used as a perturbation parameter. By changing the extra
coordinates as $\xi^{a}\longrightarrow \varepsilon^{1/2}\xi^{a}$,
the Dalambertian (\ref{eq38}) and confinemening potential can be
expanded in powers of $\varepsilon$, leading to
\begin{eqnarray}
\varepsilon \Box = \Box^{(0)} + \varepsilon \Box^{(1)} +\cdots,
\end{eqnarray}
and
\begin{eqnarray}
{\cal V}(\xi) = \frac{1}{2\varepsilon^2}\omega^2
g_{ab}\xi^{a}\xi^{b} + {\cal O}(\xi^3),
\end{eqnarray}
where
\begin{eqnarray}
\Box^{(0)} = -\frac{1}{2}g^{ab}\partial_{a}\partial_{b},
\end{eqnarray}
and
\begin{eqnarray}
\Box^{(1)} = -\frac{1}{2|\bar{g}|^{1/2}}{\cal
D}_{\mu}\bar{g}^{\mu\nu}|\bar{g}|^{1/2}{\cal D}_{\nu} +
\frac{1}{8}\bar{g}^{\mu\nu}\bar{g}^{\alpha\beta} \left(
g_{ab}\bar{K}^{a}_{\,\,\,\,\mu\nu}\bar{K}^{b}_{\,\,\,\,\alpha\beta}
-
2g_{ab}\bar{K}^{a}_{\,\,\,\,\mu\alpha}\bar{K}^{b}_{\,\,\,\,\nu\beta}
\right).
\end{eqnarray}
Thus, the $\varepsilon\rightarrow 0$ limit can be unambiguously
achieved by considering $\varepsilon{\cal H}$, so that
\begin{eqnarray}
\varepsilon{\cal H} = H_{0} + \varepsilon H, \label{eq42}
\end{eqnarray}
where
\begin{eqnarray}
H_{0}= \Box^{(0)}+ \frac{1}{2}\omega^2g_{ab}\xi^{a}\xi^{b},
\end{eqnarray}
and
\begin{eqnarray}
H = \Box^{(1)} - \frac{1}{2} M^2.
\end{eqnarray}
In the last step for obtaining an effective KG equation on the
brane, we need to ``freeze'' the extra degrees of freedom and thus
assume
\begin{eqnarray}
\Phi(x,\xi) = \Sigma_{\beta}\phi_{\beta}(x)\theta_{\beta}(\xi),
\end{eqnarray}
so that the index $\beta$ runs over any degeneracy that exists in
the spectrum of the normal degrees of freedom. The KG equation
associated with the normal degrees of freedom describes $(m-4)$
uncoupled harmonic oscillators
\begin{eqnarray}
\frac{1}{\varepsilon}\left(\Box^{(0)} + \frac{1}{2}\omega^2
g_{ab}\xi^{a}\xi^{b} \right)\theta_{\beta} = E_{0}\theta_{\beta}.
\label{eq45}
\end{eqnarray}

At this point, it would be appropriate to justify the use of the
simple harmonic oscillator eigenvalues to describe the system when
the classical potential is only taken to be parabolic near the
brane. The curvature radii of the background $V_4$ are the
$4\times m$ values $l^a_\mu$ of $\xi^a$, satisfying the
homogeneous equation \cite{eizen}
\begin{eqnarray}
\left(\bar{g}_{\mu\nu}-\xi^a\bar{K}_{\mu\nu a}\right)dx^\mu=0,
\label{eqex1}
\end{eqnarray}
where $a$ is fixed. The single scale of curvature $\rho$ is the
smallest of this solutions, corresponding to the direction in
which the brane deviates more sharply from the tangent plane.
Considering all contributions of $l^a_\mu$ in such a way that the
smaller solution of (\ref{eqex1}) prevails, the scale of curvature
may be expressed as
\begin{eqnarray}
\frac{1}{\rho}=\sqrt{\bar{g}^{\mu\nu}g_{ab}\frac{1}{l^\mu_a}\frac{1}{l^\nu_b}}.
\label{eqex2}
\end{eqnarray}
Since equation  (\ref{eq7}) can also be written as
\begin{eqnarray}
g_{\mu\nu}=\bar{g}^{\alpha\beta}\left(\bar{g}_{\mu
\alpha}-\xi^a\bar{K}_{\mu \alpha a}\right)\left(\bar{g}_{\nu
\beta}-\xi^b\bar{K}_{\nu \beta b}\right),  \label{eqex3}
\end{eqnarray}
it follows that   the above equation becomes singular at the
solutions of equation (\ref{eqex1}). Therefore, $g_{\mu\nu}$ and
consequently the metric of the bulk, described by equation
(\ref{eq6}), becomes also singular at the points determined by
those solutions. Of course this is not a real singularity but a
property of the Gaussian system. However, this singularity breaks
the continuity and regularity of the Gauss-Codazzi-Ricci equations
which are constructed with this system. Therefore, it represents
also a singularity for the wave equation of the graviton written
in the Gaussian system of the bulk. Hence the scale of curvature
$\rho$ sets a local limit for the region in the bulk accessed by
gravitons associated with those high frequency waves. Since we
have assumed $\omega/\varepsilon \gg \rho^{-2}$, the above
approximation regarding the confining potential is well justified
all the way before the singularity is reached by the extra
dimensions.  Note that changing the extra coordinates makes the
divergence in the Harmonic potential in equation (\ref{eq45}) to
disappear  and we can study the $\varepsilon\longrightarrow 0$
limit unambiguously by considering $\varepsilon H$. Equation
(\ref{eq45}) has the largest contribution of the order ${\cal
O}(1/\varepsilon)$ to the $4D$ mass of the particle. Now by
projecting the KG equation onto the space of degenerate states
$\{\theta_{\beta}\}$ the effective KG equation in $4D$ becomes
\begin{eqnarray}
\left[ -\frac{1}{|\bar{g}|^{1/2}}( \partial_{\mu}{\cal I} -i{\cal
A}_{\mu})\bar{g}^{\mu\nu}|\bar{g}|^{1/2}( \partial_{\nu}{\cal I} -
i{\cal A}_{\nu} ) + {\cal Q} - m^2 \right]\phi(x) =0. \label{eq46}
\end{eqnarray}
where
\begin{eqnarray}
\begin{array}{lll}\vspace{.5 cm}
{\cal A}_{\mu} = \frac{1}{2}A_{\mu}^{ab}\langle{i\cal L}_{ab}\rangle,\\
\vspace{.5 cm} {\cal Q} = \frac{1}{4}\left(
g^{ab}\bar{K}_{a}\bar{K}_{b} - 2\bar{K}_{\alpha\beta
a}\bar{K}^{\alpha \beta a}\right){\cal I}\\
m^2 =( M^2 + E_{0}){\cal I},
\end{array}, \label{eq47}
\end{eqnarray}
with $\langle{\cal L}_{ab}\rangle$ being the matrix obtained by
bracketing ${\cal L}_{ab}$ between the eigenstates corresponding
to $E_{0}$ and resulting in components different from zero if the
normal wavefunction lies in a degenerate, nontrivial
representation of $SO(p-1,q-3)$, and ${\cal I}$ is the $n\times n$
identity matrix ($n$ is the order of degeneracy in
equation(\ref{eq45})). In obtaining equation (\ref{eq45}) we have
used the inverse of $g_{\mu\nu}$. However, since the inverse of
$g_{\mu\nu}$ cannot be obtained in exact form, one should resort
to an expansion of its terms obtaining
\begin{eqnarray}
g^{\mu\nu} = \bar{g}^{\mu\nu} +
2\sqrt{\varepsilon}\xi^{a}\bar{K}^{\,\,\,\,\mu\nu}_{a} +
3\varepsilon\xi^{a}\xi^{b}\bar{K}^{\mu}_{\,\,\,\,\alpha
a}\bar{K}^{\,\,\,\,\nu\alpha}_{b} + {\cal O}(\varepsilon^{3/2}).
\end{eqnarray}
Equation (\ref{eq46}) shows that the extrinsic contributions stem
from the momentum independent potential ${\cal Q}$ and the
minimally coupled gauge field ${\cal A}_{\mu}$. In the classical
equation of motion these quantities do not exist and are therefore
purely quantum mechanical effects arising from higher dimensions.
On the other hand, the equation of $4D$ mass, the last equation in
(\ref{eq47}), shows that the observable mass is a quantized
quantity. If the mass of a particle in the bulk space is taken to
be zero according to the induced matter theory \cite{wesson2},
then
\begin{eqnarray}
m^2 =
\frac{\omega}{\varepsilon}\Sigma_{a}\epsilon_{a}\left(n_{a}+\frac{1}{2}\right).
\label{eq49}
\end{eqnarray}
This equation deserves a short discussion.  If the extra
coordinates are taken to be timelike then the mass will become
tachyonic. However,  if we have two extra dimensions, one being
timelike and other spacelike, then the zero point energy vanishes
and we have massless particles. This means that the fundamental
mass can be zero. In the case where they are spacelike it would be
impossible to obtain a massless particle. In equation (\ref{eq49})
as $\varepsilon\rightarrow 0$ the  mass becomes very large. On the
other hand if we change the coordinates according to
$x^\mu\rightarrow\sqrt{\varepsilon}x^\mu$, then  equation
(\ref{eq46}) gives our redefined mass
\begin{eqnarray}
\tilde{m}^2 =
\omega\Sigma_{a}\epsilon_{a}\left(n_{a}+\frac{1}{2}\right).
\label{eq50}
\end{eqnarray}

In the first section we considered $\omega$ to be of order
$\rho^{-2}$. This implies $\omega\sim\Lambda$ where $\Lambda$ is
the cosmological constant of the brane. Now, inserting the
appropriate units into equation (\ref{eq50}), we obtain the
fundamental mass $\tilde{m}_0$
\begin{eqnarray}
\tilde{m}_0\sim\frac{\hbar}{c}\Lambda^{1/2}\sim
10^{-65}\hspace{1mm} \mbox{gr}.
\end{eqnarray}
Since our change of coordinates amounted to
$x^\mu\rightarrow\sqrt{\varepsilon}x^\mu$, we relate this mass to
the micro-world. It is the mass of a quantum perturbation in a
spacetime with very small curvature measured by the astrophysical
value of $\Lambda$ as opposed to the mass sometimes inferred from
the zero point or vacuum fields of particle interactions.

As we noted above, the mass denoted by $m$ is a consequence of the
large scale  gravitational effects and becomes very  large
according to $m=\tilde{m}/\sqrt{\varepsilon}$. One may interpret
this as having a parameter relating the large scale to that of the
the small. To have a feeling of $\varepsilon$, one may use the
fundamental constants to construct it. In doing so, we note that
according to our guess represented by the relation (\ref{eq1122}),
the following combination of $\hbar$, $c$, $G$ and $\Lambda$ would
serve our purpose and renders a dimensionless quantity, {\it viz.}
\begin{eqnarray}
\varepsilon\sim\left(\frac{\hbar G\Lambda}{c^3}\right)^2\sim
10^{-240}.
\end{eqnarray}
Having made an estimate for $\varepsilon$, we can now obtain an
order of magnitude for $m_0$
\begin{eqnarray}
m_0\sim \frac{c^2}{G\Lambda^{1/2}}\sim
10^{56}\hspace{1mm}\mbox{gr}. \label{eq53}
\end{eqnarray}
This is the same as the mass of the observable part of the
universe ($10^{80}$ baryons of $10^{-24}$ gr for each)
\cite{wesson}.

Another important problem that can be addressed in the context of
the present discussion is that of the cosmological constant
\cite{sahni}. As was noted in section two, the squeezing parameter
$\varepsilon$ was introduce as a consequence of requiring a test
particle to be confined to our brane. This parameter however,
opens up an opportunity in the way of observing our universe from
two different angles, namely, either looking at our universe in
its entirety, that is, at its large scale structure  or do the
opposite, {\it i.e.} observing it from small scales. The parameter
$\varepsilon$ provides us with the tool necessary to achieve this.
It suffices to define the change of variable introduced earlier,
that is $x'^\mu=\sqrt{\varepsilon}x^\mu$. This relation connects
two very different scales; small and large. Now the question of
the disparity between the values of the cosmological constant in
cosmology and particle physics reduces to its measurement from two
different scales. If we look at it from the large scale point of
view, we measure its astrophysical value, $\Lambda\sim
10^{-56}\hspace{1mm}\mbox{cm}^{-2}$. On the other hand, if one
looks at its value from very small scales, it turns out to be
related to that of $\Lambda$ through the relation
$\tilde{\Lambda}=\Lambda/\varepsilon$ which is 240 order of
magnitude larger than its astrophysical value. The unusually large
order of magnitude should not alarm the reader for if we had used
the Plank mass in equation (\ref{eq53}) instead of $m_0$, we would
have obtained the usual value for the order of magnitude, that is
120. The above discussion leads us to the conclusion that the vast
difference between the values of the cosmological constant simply
stems from our measurements at two vastly different scales.
\section{Conclusions}
In this paper, we have presented a new model for the confinement
of test particles on the brane. In doing so, we have obtained a
confinement condition which imposes an algebraic constraint on the
extrinsic curvature. This condition is particularly helpful when
calculating the components of the extrinsic curvature. This is so
because if the metric of the bulk space is known then the
calculation of the components of the extrinsic curvature  is a
simple matter. However, when the only known quantity is the metric
of the brane, then the Codazzi equation cannot  give the
components of the extrinsic curvature uniquely  since there is
always a constant of integration. The above mentioned condition
has the advantage of being able to determine the extrinsic
curvature without ambiguity.

We also showed that a classical test particle cannot feel the
effects of the extra dimensions whereas when the system is
quantized, the effects of these  dimensions are felt by the test
particle as the gauge fields and the extrinsic potential $\cal Q$.
Also, the mass turned out to be a quantized quantity which is
related to the cosmological constant.

A major ingredient of this model is a constraining force
introduced in order to confine the test particle on the brane. We
assumed that the potential generating this force has a deep
minimum around the brane allowing an expansion of it in terms of a
parameter $\varepsilon$ which controls the size of the minimum.
This parameter plays an important role in the measurement of the
value of the cosmological constant. Since the depth of the
confining potential is a relative quantity depending on the scale
at hand, the value of the cosmological constant depends very much
on the scale defining the frame from which observations are made.
This seems to be the root of the huge disparity between the values
of the cosmological constant as measured in cosmic scales and that
which results from particle physics.


\begin{thebibliography}{99}
\bibitem{shen} Jian Qi Shen {\it Comoving suppression mechanism
and cosmological constant problem} ({\it Preprint} gr-qc/0401077)
\bibitem{horava} Horava P and Witten E 1996, {\it Eleven dimensional
supergravity on a manifold with boundary} {Nucl. Phys.} B{\bf
475}94
\bibitem{maia1}Maia M D 2002, {\it Geometry of brane-worlds}, {\it Phys. Lett.} A {\bf 297}  9
\bibitem{jaffe}Schuster P C and Jaffe R L 2003, {\it Quantum mechanics
on manifolds embedded in euclidean space}, {\it Annals Phys.}
{\bf307} 132
\bibitem{seahra} Seahra S S 2002, {\it The dynamics of test
particles and pointlike gyroscopes in the brane world and other
5$D$ models}, {\it Phys. Rev.} D {\bf 65} 124004\\
Seahra S S 2003, {\it Classical confinement of test particles in
higher-dimensional medels: stability criteria and a new energy
condition},  {\it Phys. Rev.} D {\bf 68} 104027
\bibitem{ricci}Ricci G 1903 {\it Rend. dei Lincei. ser} {\bf 5}, $12^1$ 409
\bibitem{shiro}Shiromizu T, Maeda K and Sasaki M (2000), {\it The Einstein
equations on the 3-brane world} {\it Phys. Rev.} D {\bf 62} 024012
\bibitem{baby}Binetruy P, Deffayet C and
Langlois D 2000, {\it Non-convensional cosmology from a brane
universe} {\it Nucl. Phys.} B {\bf 565} 269 ({\it Preprint}
hep-th/9905012)\\Binetruy P, Deffayet C, Ellwanger U and Langlois
D 2000, {\it Brane cosmological evolution in a bulk with
cosmological constant} {\it Phys. Lett.} B {\bf 477} 285 ({\it
Preprint} hep-th/9910219)
\bibitem{tavakol}Anderson E and Tavakol R 2003, {\it Reformulation and
interpretation of SMS braneworld} {\it Class Quantum Grav.} {\bf
20} L267 ({\it Preprint} gr-qc/0305013)
\bibitem{maia2} Maia M D, Monte E M and  Maia J F M 2004,
{\it The accelerating universe in brabe-world comology},
 {\it Phys. Lett.} B {\bf 585} 11
 \bibitem{x}Chiba T, Sugiyama N and Nakamura T 1997, {\it Cosmology with X-matter}
 {\it Mon. Not.
 R. Astron. Soc.} {\bf 289} L5\\
Chiba T, Sugiyama N and Nakamura T 1998, {\it Observational tests
of X-matter models} {\it Mon. Not.
 R. Astron. Soc.} {\bf 301} 72
 \bibitem{mit}Ikegami M, Nagaoka Y, Takagi S and Tanzawa T 1992
 {\it Prog. Theor. Phys.} {\bf 88} 229
\bibitem{eizen}Eisenhart L P, {\it Riemannian Geometry}, Princeton U. P.,
 Princeton, N. J. (1966)\\
Aminov Y, {\it The Geometry of Submanifolds}, Gordon and Breach,
Science Publishers (2001).
\bibitem{wesson2}Wesson P S, {Space-Time-Matter}: {\it Modern
Kaluza-Klein Theory}, World Scientific, Singapore, (1999)
\bibitem{wesson}Wesson P S, {\it Is mass quantized}
({\it Preprint} gr-qc/0309100)\\
Mongan T R, {\it Vacuum dominance and holography} ({\it Preprint} gr-qc/0310015)\\
Carneiro S and Lima J A S, {\it Decaying $\Lambda$ cosmology,
varying $G$ and holography}  ({\it Preprint} gr-qc/0405141)
\bibitem{sahni} Sahni V and Starobinsky A, {\it The case for a positive
cosmological $\Lambda$-term}, 2000 {\it Int. J. Mod. Phys.} D {\bf
9} 373\\
Peebles P J E and Ratra B 2003, {\it The cosmological constant and
dark energy}, {\it Rev. Mod. Phys.} {\bf 75} 559\\
Padmanabhan T 2003, {\it Cosmological constant - the weight of the
vacuum}, {\it Phys. Rept.} {\bf 380}  235


\end{thebibliography}
\end{document}